\documentclass[useAMS,usenatbib]{mn2e}
\usepackage{graphics,epsfig,txfonts,color}
\usepackage{amssymb}
\usepackage{url}

\usepackage{color}
\usepackage{natbib}
\usepackage{tabularx}

\newcommand{\HESSJ}{HESS~J1507-622}

\def\degr{\hbox{$^\circ$}}
\def\arcmin{\hbox{$^\prime$}}

\title[Gamma-ray sources along the super-galactic plane]{Is there a population of unidentified gamma-ray sources distributed along the
super-galactic plane?}

\author[W. Domainko]
  {Wilfried F. Domainko \\
    Max-Planck-Institut f\"ur Kernphysik, P.O. Box 103980, D-69029 Heidelberg, Germany}

\date{}

\pagerange{\pageref{firstpage}--\pageref{lastpage}} \pubyear{20xx}

\def\LaTeX{L\kern-.36em\raise.3ex\hbox{a}\kern-.15em
    T\kern-.1667em\lower.7ex\hbox{E}\kern-.125emX}

\begin{document}

\label{firstpage}

\maketitle

\begin{abstract}
The distribution on the sky of unidentified sources at the highest energies
where such a population is evident is investigated.
For this purpose, sources without identification in the first Fermi-LAT catalog $>$10\,GeV (1FHL) that
are good candidates for detection above the 50-100~GeV regime are selected.
The distributions of these objects around the Galactic and super-galactic plane are explored.
By using a Kolmogorov-Smirnov test it is examined if these sources are distributed
homogeneously around these planes. Surprisingly, an indication for 
an inhomogeneous distribution is found for the case of the super-galactic plane
where a homogeneous distribution can be excluded by a confidence level of 95\%.
On a 90\% confidence level also a homogeneous distribution of sources around
the Galactic plane can be excluded.
For the hypothesis that this reflects the true distribution of sources 
rather than a statistical fluctuation,
implications for the underlying source populations are discussed.
\end{abstract}


\section{Introduction}

The distribution on the sky is a powerful tool to explore the nature and distance scale
of unidentified astrophysical sources. It has already been applied to steady 
\citep[e.g.][]{gehrels2000,aharonian2006} and transient sources \citep[e.g.][]{meegan1992,foley2008,thornton2013}.
Alignment with prominent features on the sky like the Galactic plane or local galaxies or galaxy clusters
may connect the sources to these structures. A homogeneous distribution may either point
to an origin in the solar neighborhood on the one extreme of possible distance scales or to a 
cosmological origin on the other extreme. 

The aim of this work is to investigate the distribution on the sky of unidentified sources at the highest energies
where such a population is evident.
Several unidentified sources radiating in the energy regime between 10-100\,GeV have been reported
from the \emph{Fermi} satellite \citep[The first Fermi-LAT catalog of $>$10\,GeV sources (1FHL)][]{ackermann2013}.
A subsample of the 1FHL sources has been flagged for being good candidates
for detection above the 50-100~GeV regime. Sources selected in this way are here used as a proxy for the distribution of such sources
in the very-high energy (VHE, E$>$100\,GeV) gamma-ray band. The distribution on the sky of these
sources is investigated in this work.

H.E.S.S. has discovered an unidentified VHE gamma-ray source
with a significant off-set from the Galactic plane: \HESSJ\ \citep{acero2011}. Despite significant observational
and theoretical effort the nature of the source still remains elusive \citep{domainko2011,domainko2012,vorster2013,tibolla2014,eger2014}.
Sources that are good candidates for detection in the 50-100~GeV regime in the 1FHL catalog appear similar in their GeV properties to 
1FHL~J1507.0-6223, the \emph{Fermi} counterpart of \HESSJ\ .
So far the relation between these sources and \HESSJ\ is not known.
The \emph{Fermi} counterpart of this source 
has been reported as being point-like \citep{acero2013}, however, it has to be noted, that \HESSJ\ is clearly extended in the VHE regime. 
In this paper also the implications for the case are discussed where \HESSJ\ is a representative
of a source class that is similarly distributed on the sky as the 1FHL sources that are good
candidates for detection above the 50-100~GeV regime.

\section{Sample selection}
\label{sec:sample}

For the purpose of this work, the subsample of the 1FHL that are 
good candidates for detection at energies above 50-100~GeV and lack an identification is selected as source sample.
The advantage of using the 1FHL catalog is the fact that it provides a homogeneous 
sky coverage.
None of these sources has been flagged as being variable.

Good candidates for detection above 50-100\,GeV may in principle belong to a similar object type as \HESSJ\ .
Contrary to \HESSJ\ all unidentified sources detected in the VHE gamma-ray
range with H.E.S.S. are located very close to the Galactic plane \citep[within $\pm 1$\degr of the Galactic plane, see][]{aharonian2008}. 
At this point it is not clear if \HESSJ\ is a representative of this source population
or if it is a member of a different type of objects with a different distribution on the sky 
\citep[see also][for a discussion]{eger2014}. Here the second possibility is explored.

It is investigated whether there is an indication for a
population of sources, that may belong to an old stellar population (as indicated by their off-set from the Galactic plane)
and whether there is any indication for a non-uniform distribution on the extragalactic sky.
Therefore, the region of $\pm 1$\degr around the Galactic disk is excluded from the selection of the source sample.
These sources are very likely of galactic
origin and most likely connected to a young stellar population. 
This results in a selection of 29 sources (see Tab. \ref{tab_sources}).
For a homogeneously distributed
sample of 29 sources on the sky only 0.5 sources would be expected within $\pm 1$\degr of the Galactic plane.

\begin{table}
\begin{center}
\caption[]{Unidentified sources from the 1FHL catalog that are good candidates for
detection above the 50-100~GeV regime. The region of $\pm 1$\degr around the Galactic disk is excluded.}
\begin{tabular}{l|rrrr}
\hline
\hline
1FHL name & SGL & SGB & l & b \\
\hline
J0030.1-1647 & 278.6 & 1.2    &   96.3 & -78.6 \\
J0053.9+4030 & 335.8 & 10.3   &  123.4 & -22.4 \\
J0110.0-4023 & 257.7 & -12.9  &  287.9 & -76.2 \\
J0307.4+4915 & 353.0 & -7.9   &  144.6 & -7.8 \\
J0312.8+2013 & 328.1 & -24.7  &  62.5  & -31.6 \\
J0338.4+1304 & 324.4 & -33.5  &  73.5  & -32.9 \\
J0425.3+6320 &  10.5 & -5.8   &  144.4 &   9.8 \\
J0425.4+5601 &  6.2  & -11.7  &  149.7 &   4.7 \\
J0432.2+5555 &  6.9  & -12.4  &  150.5 &   5.3 \\
J0439.9-1858 & 285.7 & -57.7  &  216.9 &  -37.2 \\
J0509.9-6419 & 220.6 & -38.2  &  274.3 &  -35.2 \\
J0601.0+3838 &  13.7 & -34.2  &  173.2 &  7.6 \\
J0639.6-1244 & 335.0 & -85.3  &  223.0 &  -8.3 \\
J0650.4+2056 & 24.8  & -53.3  &  194.0 &  9.2 \\
J0746.3-0225 & 71.3  & -71.7  &   221.5 & 11.1 \\
J0803.4-0334 & 82.5  & -69.3  &   224.6 & 14.2 \\
J1115.0-0701 & 116.7 & -25.9  &   265.1 & 48.6 \\
J1129.2-7759 & 194.6 & -19.8  &  298.6  & -15.8 \\
J1240.4-7150 & 187.5 & -16.1  &  302.1  &  -9.0 \\
J1315.7-0730 & 125.1 &  2.9   &  313.4  & 54.9 \\
J1328.5-4728 & 164.2 & -6.0   &   309.4 & 14.9 \\
J1353.0-6642 & 183.6 & -8.8   &  309.0  & -4.6 \\
J1406.4+1646 & 104.2 & 21.1   &  6.0  &  69.8 \\
J1440.6-3847 & 161.4 & 9.4    &  325.2 &  19.3 \\
J1507.0-6223 & 183.6 & 0.2    &  317.9 & -3.5 \\
J1545.2-6640 & 189.5 & 0.5    &  319.0 & -9.3 \\
J2004.7+7003 &  19.3 & 34.6   &  102.9 & 19.5 \\
J2036.9-3325 & 232.5 & 35.2   &  9.7   & -35.5 \\
J2159.1-3344 & 247.4 & 23.9   &  12.0  & -52.6 \\
\hline
\end{tabular}
\label{tab_sources}
\end{center}
\end{table}

\section{Distribution of the sources}

\HESSJ\ is located close to both the Galactic and the super-galactic plane 
$(b \approx -3.5\degr, SGB \approx 0.2\degr)$. In the following
it is investigated how the selected source sample is distributed with respect to these
two planes. In all cases a Kolmogorov-Smirnov (KS) test is used to check if the sources
are homogeneously located around a given plane.

\subsection{Galactic plane}
\label{sec:gal}

\begin{figure}
\centering
\includegraphics[height=6.2cm]{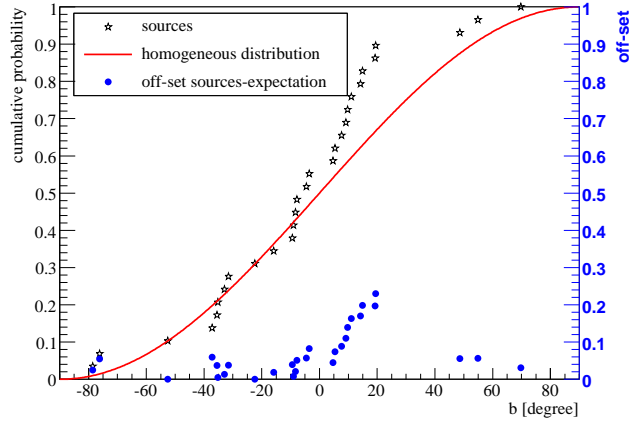}
\caption{Kolmogorov-Smirnov test for the distribution of the sources of Tab. \ref{tab_sources} around the Galactic plane.
The largest off-set from a homogeneous distribution $D_{max}$ is 0.230.}
\label{figure:gal}
\end{figure}

The KS test has been applied to test if the selected sample of sources is distributed
homogeneously around the Galactic plane (see Fig. \ref{figure:gal}). As a result of this test the maximum off-set
$D_{max}$ of the cumulative probability of the sources from the prediction for a homogeneous
distribution is 0.230 for 29 sources. Therefore, the hypothesis of a homogeneous
distribution can be rejected by a confidence level of 90\%. 

\subsection{Super-galactic plane}
\label{sec:super} 

The super-galactic plane is the plane on the sky where galaxies with distance up to a few
10s~of~Mpc cluster \citep{devacouleurs1976,lahav2000}.
The same procedure as in Sec. \ref{sec:gal} has been applied for the distribution of sources
around the super-galactic plane. There $D_{max}$ is found to be 0.273. Consequently,
the hypothesis of a homogeneous
distribution can be rejected at a confidence level of 95\%.

\begin{figure}
\centering
\includegraphics[height=6.2cm]{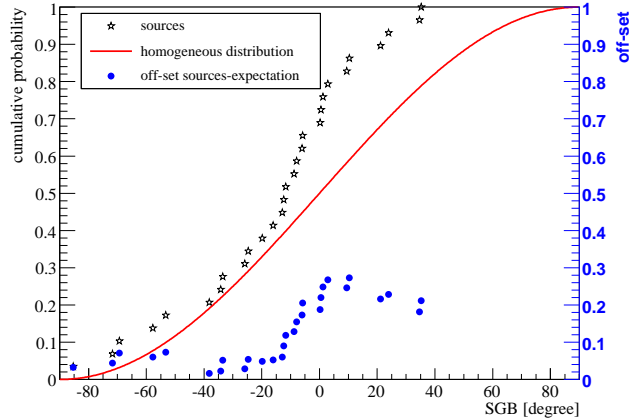}
\caption{Same as Fig. \ref{figure:gal} but for the super-galactic plane.
$D_{max}$ is 0.273.}
\label{figure:super}
\end{figure}

\subsection{Discussion}

Clustering of sources around the Galactic plane is expected due to the presence of Galactic objects
whereas a concentration of unidentified sources around the super-galactic plane is rather unexpected.
Therefore, here it is investigated if the apparent inhomogeneous distribution of sources 
is artificially generated by potential inhomogeneous
detection efficiency by \emph{Fermi} across the sky.

\begin{figure}
\centering
\includegraphics[height=6.2cm]{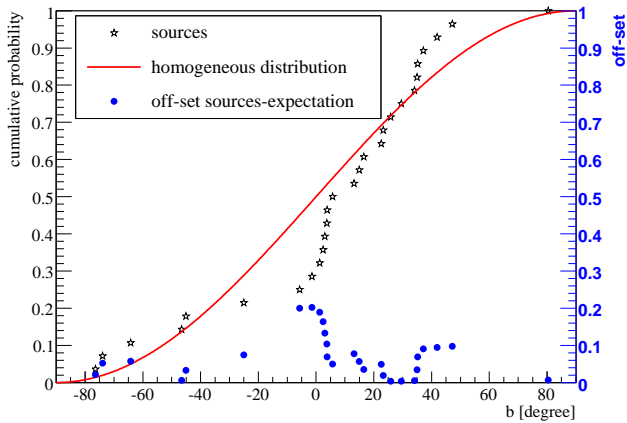}
\caption{KS test for the distribution of unidentified sources from the 1FHL catalog
that are not flagged for being good candidates for detection above 50-100~GeV around the Galactic plane.
$D_{max}$ is 0.202.}
\label{figure:lowgal}
\end{figure}

\begin{figure}
\centering
\includegraphics[height=6.2cm]{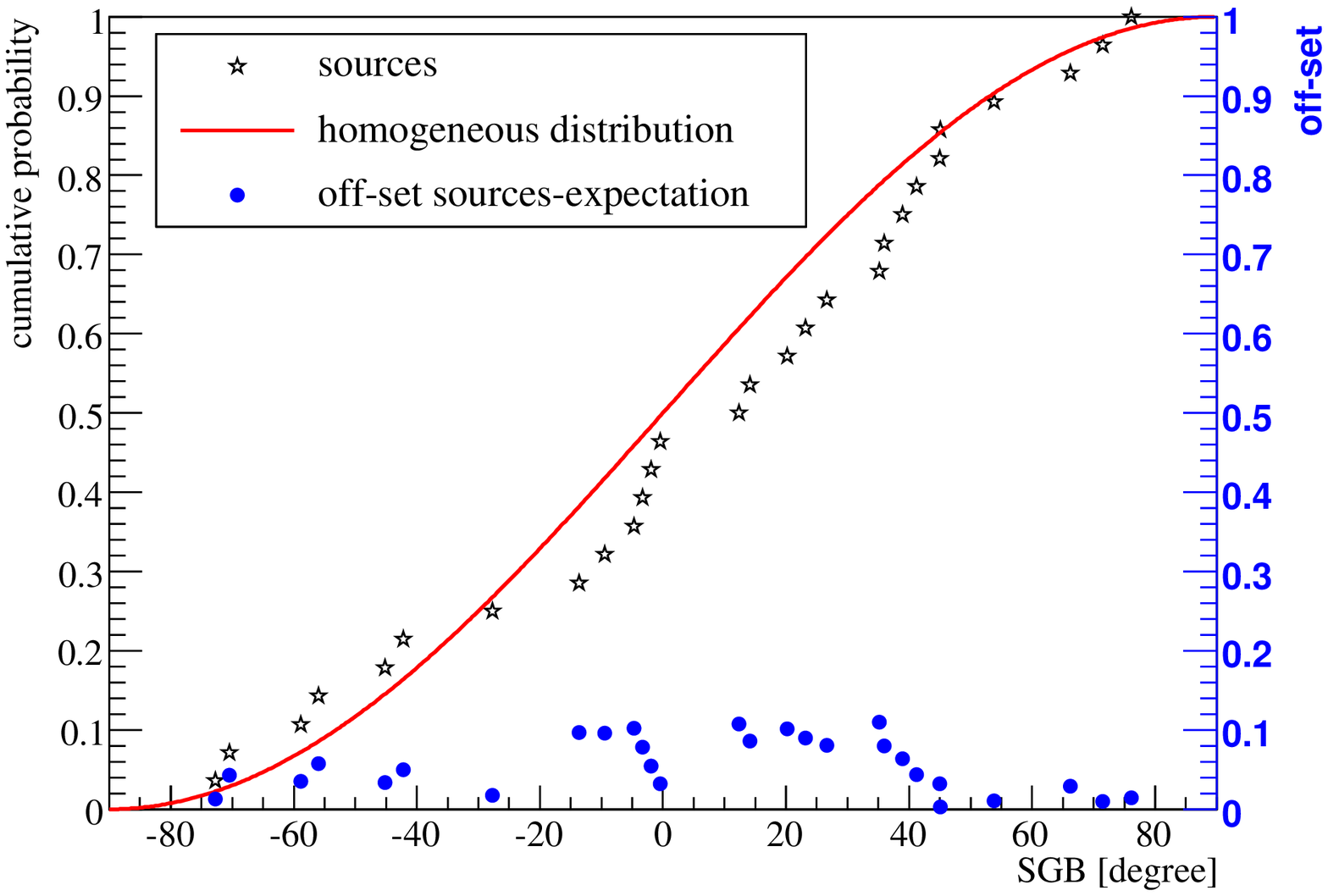}
\caption{Same as Fig. \ref{figure:lowgal} but for the super-galactic plane.
$D_{max}$ is 0.110.}
\label{figure:lowsuper}
\end{figure}

From the sample of selected sources it is evident, that there is a deficit of
sources in the direction of the super-galactic north-pole.
The super-galactic north pole is located in Galactic coordinates in the
direction of l = 47.37\degr\ and b = +6.32\degr\ and thus rather close to the Galactic plane
\citep{devacouleurs1976,lahav2000}.
Two effects may be important there.
On the one hand source detection might be affected by the presence of Galactic diffuse emission.
On the other hand, since there is some indication of sources clustering around the Galactic plane,
it could actually be expected to find an excess of sources at low Galactic latitudes (and also in 
the direction of the super-galactic north pole).
These two effects work in opposite directions.
To test if there is any bias of this kind, the same procedure as in Sec. \ref{sec:gal} and \ref{sec:super} 
is applied to unidentified sources in the 1FHL that were not flagged to be good candidates for detection
above 50-100~GeV. 
It would be expected that detection efficiency for theses sources is biased in a comparable way
as for the sample of sources defined in Sec. \ref{sec:sample}.
Again, sources with Galactic latitudes in between $\pm 1$\degr\ were excluded
(see Sec. \ref{sec:sample}). This procedure selects a sample of 28 sources. The result of a KS test for these
sources is shown in Fig. \ref{figure:lowgal} (for the Galactic plane) and Fig. \ref{figure:lowsuper} 
(for the super-galactic plane). The $D_{max}$ found for this sample of sources is 0.202 (Galactic plane) and 0.110 (super-galactic plane).
As a result, the location of these sources with respect to the super-galactic plane is compatible with a homogeneous distribution.
Therefore, it is unlikely that the indication of an inhomogeneous distribution found for the sample of sources 
given in Tab. \ref{tab_sources} is introduced by an inhomogeneous detection efficiency of \emph{Fermi}.

\begin{figure}
\centering
\includegraphics[height=6.6cm]{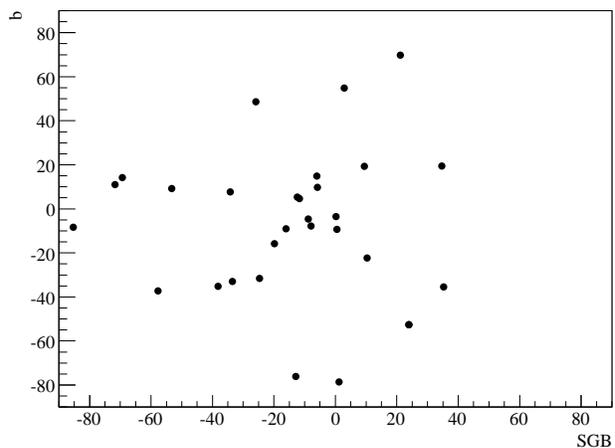}
\caption{Correlation between the Galactic and the super-galactic latitude of the 
sources from Tab. \ref{tab_sources}. No clear correlation between these two quantities
is seen.}
\label{figure:corr}
\end{figure}

Additionally, it is investigated if there is any correlation between
the Galactic latitude and the super-galactic latitude of the sources in Fig. \ref{figure:corr}.
As a result, no evidence of an obvious correlation is found.
The reason for the deficit of sources in the direction of the super-galactic north-pole is
not know at the moment.

Finally, the same procedure as in Sec. \ref{sec:gal} and \ref{sec:super} is applied to the sample of 36
high-energy neutrino events detected by IceCube \citep{aartsen2014}. As a result of this investigation a $D_{max}$ of
0.143 (Galactic plane) and 0.103 (super-galactic plane) is found and thus the distribution
of events is consistent with a homogeneous distribution for both cases.

To conclude, there are moderately significant indications that sources from the sample defined in
Sec. \ref{sec:sample} are inhomogeneously distributed with respect to both, the Galactic and the
super-galactic plane. Rather surprisingly, the strongest indication for an inhomogeneous distribution
is found for the case of the super-galactic plane for the sources of Tab. \ref{tab_sources}.

\section{Implications}

For this Section it is assumed that the inhomogeneous distribution of
sources with respect to the Galactic and super-galactic plane is not a statistical
fluctuation but rather reflects the true distribution of sources.

\subsection{Galactic origin}

The discussion in this section is based on the hypothesis that some of the sources
located in the vicinity of the Galactic plane from the sample defined in Sec. \ref{sec:sample} 
are of similar origin as HESS~J1507-622. It has to be noted that at the moment its not clear
whether such a connection exists. The extended nature of \HESSJ\ in the VHE gamma-ray range would favor
a Galactic origin.

\HESSJ\ is likely located at multiple-kpc distance as indicated by its compact appearance \citep{acero2011,domainko2012}
and its considerably absorbed potential X-ray counterpart \citep{eger2014}. For the following discussion
it is assumed that some of the sources in the sample defined in Sec. \ref{sec:sample} are located at a comparable distance.
The typical flux above 10\,GeV of theses sources is about $5 \times 10^{-12}$~erg~cm$^{-2}$~s$^{-1}$ \citep{ackermann2013},
resulting in a luminosity of about $10^{34}(d/5\, \mathrm{kpc})^2$~erg\,s$^{-1}$,
with $d$ being the typical distance to these objects.
Assuming IC up-scattering of CMB photons by highly energetic electrons as the dominating radiation mechanism 
\citep[cooling time of electrons with energy
of 300\,GeV of about $3 \times 10^6$~years,][]{hinton2009} then the energy in electrons would be about $10^{48}(d/5\, \mathrm{kpc})^2$~erg.
The off-set for sources located close to the Galactic plane in Tab. \ref{tab_sources}
is typically a few degrees. For a multi-kpc distance this would result in a Galactic scale hight of these
sources of several hundred parsecs. To summarize, the energetics of the sources would be in line with
the typical energetics of known types of VHE gamma-ray emitters \citep{hinton2009} whereas the Galactic scale hight would be much larger
than the typical scale hight of a young stellar population \citep[see e.g.][]{gregoire2013,domainko2014}. 

Finally, if it is assumed that $O (10)$ such sources exist in the Galaxy and their typical age is
about $10^6$\,years \citep[see][for a discussion on \HESSJ\ ]{eger2014}, then the typical rate
for their formation would be $10^{-5}$~yr$^{-1}$ for the Galaxy or 100~Gpc$^{-3}$\,yr$^{-1}$
\citep[assuming a density of Milky Way-type galaxies of 0.01~Mpc$^{-1}$,][]{cole2001}.

\subsection{Extragalactic origin}

Firstly, and more generally, here the implications of a source population connected to the
super-galactic plane is explored. The typical distance of galaxies that form the super-galactic plane
is a few tens of Mpc \citep{lahav2000}. 
Therefore, for $O (10)$ such sources their volume density would be roughly $4 \times 10^{-5}(d/30\, \mathrm{Mpc})^{-3}$~Mpc$^{-3}$.
Such a distance scale would result in a typical un-beamed source luminosity of
$5 \times 10^{41}(d/30\, \mathrm{Mpc})^2$~erg\,s$^{-1}$ and typical energy in electrons of $5 \times 10^{55}(d/30\, \mathrm{Mpc})^2$~erg
if again it is assumed that IC up-scattering of CMB photons by highly energetic electrons is the dominating radiation process.
These energy requirements are very challenging for galactic type sources but can easily be accommodated by active galactic nuclei. 
In principle the energetics would be reduced if the
emission is significantly beamed (as it is the case for example for BL~Lac type objects).

In this extragalactic scenario it is interesting to investigate if the sources from Tab. \ref{tab_sources} are located at the position of nearby galaxies.
This search has been conducted by using the NASA/IPAC extragalactic database\footnote{http://ned.ipac.caltech.edu/}.
Only for one case (1FHL~J2159.1-3344) a galaxy with a distance of less than 50~Mpc 
\citep[GALEXASC J215915.26-335251.9, angular distance of 8.4\arcmin\, $z=0.0097$, $D \approx 40$~Mpc,][]{jones2009} has been found within 10\arcmin\
of the source. If the distance is increased to 100~Mpc, galaxies inside an angular radius of
10\arcmin\ for two more sources are found \citep[1FHL~J0053.9+4030, CGCG~536-006, angular distance 5.5\arcmin\, $z=0.019$, $D \approx 80$~Mpc,][]{strauss1992}
and \citep[1FHL~J1406.4+1646, SDSS~J140701.68+164232.8, angular distance 9.1\arcmin\, $z=0.016$, $D \approx 67$~Mpc,][]{sanchez2011}.
As a result, it appears, that there is no strong correlation between the sources from Tab. \ref{tab_sources}
and nearby galaxies.

Secondly, and more specifically, the implications of a population of extragalactic sources with similar 
properties as \HESSJ\ is investigated. An extragalactic scenario may in principle be 
motivated by the large required distance to the object if the energy density 
in particles and in the magnetic field are roughly equal \citep{domainko2014}. 
The spectral parameters found by 
\citet[][energy in electrons $5\times 10^{47} (d/1\, \mathrm{kpc})^2$~erg, magnetic field 0.47~$\mu$G]{eger2014}
would result in an equipartition distance of about 30~Mpc, that is coarsely compatible to the
typical distance of galaxies connected to the super-galactic plane.
For a source age of $10^6$~years \citep[see again][]{eger2014}, the rate of formation of such objects would roughly be 
$0.4 \times (d/30\, \mathrm{Mpc})^{-3}$~Gpc$^{-3}$\,yr$^{-1}$.
The caveat for an extragalactic scenario for \HESSJ\ is the fact that the angular extension of the source 
would require a physical extension of several tens of kpc. That is very challenging for a leptonic scenario 
\citep[see][for a discussion]{domainko2014}.

\section{Outlook}

At this point inhomogeneities in the distribution of high-energy gamma-ray sources are found with moderate
significance and its not clear whether this represents a statistical fluctuation or a real signal. 
One way to improve this situation is to perform multi-wavelength studies on some specific sources.
To test the galactic scenario, sources that are located close to the Galactic plane are of enhanced interest. 
Good candidates would be: 1FHL~J0425.4+5601 $(b \approx 4.7\degr)$ and 1FHL~J1353.0-6642 $(b \approx -4.6\degr)$. To test the
extragalactic scenario, sources that are located close to the super-galactic plane but are far from the Galactic
plane may be important. These sources would be: 1FHL~J0030.1-1647 $(b \approx -78.6\degr, SGB \approx 1.2\degr)$ and 1FHL~J1315.7-0730
$(b \approx 54.9\degr, SGB \approx 2.9\degr)$. Follow-up observations of this kind may reveal if there is any source population
connected to the Galactic and super-galactic plane.

\label{lastpage}

\end{document}